\documentstyle[12pt]{article}

\newcommand\ba{\begin{array}}
\newcommand\ea{\end{array}}
\newcommand\ben{\begin{equation}}
\newcommand\een{\end{equation}}
\newcommand\bea{\begin{eqnarray}}
\newcommand\eea{\end{eqnarray}}
\def\PR{Phys. Rev.}
\def\PRL{Phys. Rev. Lett.}

\def\APJ{Ap. J.}

\def\ga{\gamma}
\def\de{\delta}

\def\et{\eta}

\def\rh{\rho}
\def\si{\sigma}
\def\ta{\tau}

\def\Th{\Theta}

\def\Up{\Upsilon}

\def\Om{\Omega}

\def\rf#1#2#3#4#5{#1, {#3},  {#4} (#2) #5}
\def\rfbk#1#2#3#4#5{#1, #3, #5, #4 (#2)}

\def\vev#1{\langle#1\rangle}
\def\da{\dot a}
\def\pa{\partial}
\def\Xd{\dot{X}}
\def\Xp{\mathaccent 19 X}
\def\bX{\mbox{\boldmath $X$}}
\def\bx{\mbox{\boldmath $x$}}
\def\bk{\mbox{\boldmath $k$}}

\def\({\left(} \def\){\right)}
\def\[{\left[} \def\]{\right]}
\def\G{{\cal G}}
\def\ded{\dot\de}

\title{
{\normalsize
\begin{flushright}
SUSX--TP--94--75\\
December 1994\\
\end{flushright}}
\vspace{1 cm}
Cosmological Perturbations from Cosmic Strings
\thanks{From: Proceedings of `Trends in Astroparticle
Physics', Stockholm, Sweden 22-25 September 1994, edited by
L. Bergstr\"om, P. Carlson, P.O. Hulth and H. Snellman (to be published
in Nucl.~Phys~B, Proceedings Supplements Section)}
}
\author{M. Hindmarsh}

\date{}
\begin{document}
\maketitle

\begin{center}
{
\normalsize\it
School of Mathematical and Physical Sciences\\
University of Sussex\\
Brighton BN1 9QH\\
UK\\
}
\end{center}
\vfill
\begin{abstract}
Some aspects of the theory of cosmological perturbations
from cosmic strings and other topological defects are outlined,
with particular reference to a simple example:  a spatially flat
CDM-dominated universe.  The conserved energy-momentum
pseudo-tensor is introduced, and the equation for the density
perturbation derived from it.  It is shown how the scaling
hypothesis for defect evolution results in a Harrison-Zel'dovich
spectrum for wavelengths well inside the horizon.

\end{abstract}

\newpage

One of the most exciting problems in cosmology today is to explain the
origin of the primordial fluctuations, whose imprint we have seen
in the temperature of the cosmic microwave
background \cite{CMB}.  It is generally thought that these fluctuations owe
their origin to processes happening very early on in the history of the
Universe, perhaps at about $10^{-36}$ seconds after the Big Bang.  In
the last 15 years or so, two classes of theory have emerged to explain
how this came about: those based on inflation and those on
ordering dynamics after a cosmological phase transition (sometimes
loosely called topological defects).

Inflationary fluctuations (see e.g.~\cite{Inf}) arise
from the amplification of quantum fluctuations in a weakly-coupled
scalar field by the strong gravitational field of an early phase of
accelerated expansion.  They have the statistical properties of a
Gaussian random field, for which the calculations of correlation
functions (such as the power spectrum of density fluctuations) are
relatively easy.

Cosmic strings \cite{HinKib94} and other defects \cite{VilShe94},
on the other hand, are certainly not Gaussian.  They are
formed in an early phase transition, in an initial distribution more
akin to a Poisson process, although one of the major
problems of the string scenario is
to understand the statistical properties of the evolving network.  In
this talk I shall show how an understanding of string statistics can be
converted into knowledge of the statistics of the primordial
fluctuations, using the general relativistic theory of cosmological
perturbations from ``stiff'' sources developed by Veeraraghavan and
Stebbins \cite{VeeSte90}.  To bring out the essentials I
shall consider a very simple case: an $\Om=1$ universe dominated by cold
dark matter (CDM), and study the density perturbation only.  Much
of what I say applies to textures \cite{PenSpeTur94} as well.

Firstly, I outline notation and conventions.  For the space-time metric
I write
\ben
g_{\mu\nu}(\et,\bx) = a^2(\et)(\et_{\mu\nu} + h_{\mu\nu}(\et,\bx)),
\een
so that $\et$ is conformal time, $\bx$ are the comoving coordinates,
$\et_{\mu\nu} = {\rm diag}(-1,1,1,1,)$ is the Minkowksi metric, and
$h_{\mu\nu}$ is the (scaled) metric perturbation.  I shall use the
synchronous gauge $h_{0\mu} = 0$. Greek indices from near the middle
of the alphabet run from 0 to 3, while Roman ones are purely spatial
indices. Natural units are used throughout.

Strings are stiff sources: that is, their energy-momentum tensor is
separately conserved, so that
\bea
\dot\Th_{00} + \frac{\da}{a}\Th_+ + \pa_i\Th_{0i} &=& 0, \\
\dot\Th_{0i} + \frac{\da}{a}\Th_{0i} + \pa_j\Th_{ij} &=& 0.
\label{DefEOM}
\eea
These equations follow from the equations of motion of the sources,
which are unaffected by small variations in the background gravitational
field.  For strings, $\Th_{\mu\nu}$ can be expressed in terms of the
linear mass density $\mu$ and the coordinates of the string
world-sheet, $X^\mu(\si,\et)$:
\ben
\Th_{\mu\nu}(\et,\bx) =
\mu\int d\si\,(\Xd^\mu\Xd^\nu - \Xp^\mu\Xp^\nu)\de^{(3)}(\bx - \bX(\si,\et)),
\label{StrEMTen}
\een
where $\Xp^\mu = \pa X^\mu/\pa \si$ and $\Xd^\mu = \pa X^\mu/\pa\et$.

The evolution of the perturbations to the metric and to the
energy-momentum tensor of the matter content of the Universe are then
described by the linearised Einstein equations \cite{Wei72},
together with the covariant
energy-momentum conservation of the matter. This evolution
can be neatly expressed as
the {\it ordinary} conservation of an energy-momentum pseudo-tensor
$\ta_{\mu\nu}$ \cite{VeeSte90}.  For CDM plus strings (or other stiff
sources) it takes the form
\bea
\ta^{00} &=& a^2 \rh\de - {1\over 8\pi G} \frac{\da}{a} \dot
h + \Th_{00} \nonumber\\
\ta^{0i} &=&  - \Th_{0i} \nonumber\\
\ta^{ij} &=& -\frac{1}{8\pi G} \frac{\da}{a} \(\tilde{h}_{ij} -\frac{2}{3} \dot
h \de_{ij}\) + \Th_{ij},
\eea
where $\rh$ is the background CDM density, $\de$ is the fractional
perturbation to it, and $\tilde{h}_{ij}$ is the traceless part of $h_{ij}$.
The conservation equation $\pa_\mu\ta^{\mu\nu} = 0$
reduces in this case, after the use of the equations of motion of the
defects (\ref{DefEOM}) and the constraint equation $\dot h = -2\dot \de$,
to
\ben
\ddot \de + \frac{\da}{a} \dot\de - \frac{3}{2}\(\frac{\da}{a}\)^2 \de =
4\pi G\Th_+,
\label{EOM}
\een
where $\Th_+ = \Th_{00} + \Th_{ii}$.  This is a second order
inhomogeneous linear ODE, and is easily solved in terms of two Green's
functions,
\bea
\G_1(\et,\et')&=& {1\over 5}\(3\frac{\et^2}{{\et'}^2} + 2
\frac{{\et'}^3}{\et^3}\), \\
\G_2(\et,\et') &=& \frac{1}{5} \(
\frac{\et^2}{\et'} - \frac{{\et'}^4}{\et^3}\),
\eea
given a source function $\Th_+(\et,\bx)$ and initial conditions $\de_i$
and $\ded_i$ at time $\et_i$.  The initial conditions are not,
arbitrary, for they must have arisen from a causal evolution, satisfying
pseudo-energy-momentum conservation.  On super-horizon scales at
$\et_i$,  the total pseudo-energy vanishes, or
\ben
a_i^2\rh_i\de_i(\bk) + \frac{1}{4\pi G} \frac{\da}{a} \ded_i +
\Th_{00} = 0.
\label{Comp}
\een
This equation expresses the fact that energy is locally conserved when
making strings (or other defects).  The
initial energy in the strings is {\it compensated\/} by that in the
matter.

The general solution of equation (\ref{EOM}) can be split into
two parts, as is usual with inhomogeneous differential equations,
\bea
\de^I(\et) &=&  \G_1(\et,\et_i) \de_i + \G_2(\et,\et_i)\ded_i, \\
\de^S(\et) &=& \int_{\et_i}^\et d\et'\,\G_2(\et,\et')\Th_+(\et').
\eea
These were termed the {\it initial\/} and {\it subsequent\/} compensation by
Veeraraghavan and Stebbins \cite{VeeSte90}.   The initial compensation
soon settles into a purely growing mode, with
\bea
\de^I(\et) &= &\frac{1}{10}\(\frac{\et}{\et_i}\)^2
(3\de_i + \et_i\ded_i) \nonumber\\
&=& -\frac{3}{5}\(\frac{\et}{\et_i}\)^2
\frac{\Th_{00}(\et_i)}{a_i\rh_i},
\eea
where I have used (\ref{Comp}) in the second equality.  We see that the
initial compensation is entirely determined by the initial defect
density.

This raises an important question: how does the Harrison-Zel'dovich
spectrum $\vev{|\de|^2} \sim \et^4 k$ arise?  This spectrum
is special because it is scale invariant: the r.m.s.~mass fluctuations
in a horizon-sized volume are constant.  To demonstrate how a
scale-invariant spectrum emerges from the assumption of
scaling in the network, we must first show that the
initial compensation is cancelled by a piece of
the subsequent compensation.  After all, we should not expect the power
spectrum
to depend on the initial conditions, for they may well produce white noise
($k^0$), having been generated by a causal process inside the
horizon at $\et_i$. Using the equation of motion of
the defect network (\ref{DefEOM}), we find
\bea
\de^S(\et) &=& 4\pi G \[ \frac{1}{10}\({\et^2}
\frac{\et_i^5}{\et^3}\) \Th_{00}(\et_i)
- \frac{1}{2}\int_{\et_i}^\et d\et' \frac{{\et'}^4}{\et^3}
\Th_{00}(\et') \right. \nonumber\\
&+& \left. \frac{1}{10}\int_{\et_i}^\et d\et'
\({\et^2} - \frac{{\et'}^5}{\et^3}\) \Th_{0i}(\et') \].
\eea
The first term on the first line cancels the initial compensation,
the second is a
transient, and we are left with the second line as the true source of the
density perturbations in the dark matter component.  Thus it is extremely
important to get the initial conditions right in numerical
simulations, otherwise one may well find spurious white noise in
the final power spectrum, which will dominate any Harrison-Zel'dovich
component at late times \cite{PenSpeTur94}.

To calculate the power spectrum, we need the density perturbations well
inside the horizon, for which $k\et \gg 1$.  All that survives (provided
$\Th_{0\mu}(\et,\bk)$ drops off fast enough) is the growing mode, which
is the piece proportional to $\et^2$.  Thus
\ben
\vev{|\de(\et)|^2} \to \(\frac{4\pi G}{10}\)^2\et^4 k_ik_j
\int_{\et_i}^\et d\et_1 d\et_2\,
\vev{\Th_{0i}(\et_1,\bk) \Th_{0i}^*(\et_2,\bk)}.
\een
The angle brackets denote an average over an ensemble of realisations
of the defects.  We can deduce the form of
the right hand side, using the scaling hypothesis
for defects.

Firstly, we do not expect the momenta $\Th_{0i}(\et,\bk)$ to be
correlated at two different times $\et_1$, $\et_2$ if $|\et_2 - \et_1|
\gg k^{-1}$.  Let us denote the correlation time by $\Up k^{-1}$.
Secondly, statistical isotropy in the defect network means that
$\vev{\Th_{0i}(\et)\Th_{0j}^*(\et)} = {\textstyle 1 \over 3} \de_{ij}
\vev{|\Th_{0i}(\et)|^2}$.  Thus,
\ben
\vev{|\de(\et)|^2} \to \frac{1}{3}\(\frac{4\pi G}{10}\)^2\et^4 \Up k
\int_{\et_i}^{\et} d \et_1\, \vev{|\Th_{0i}(\et_1,\bk)|^2}.
\een
Thirdly, the scale invariance tells us that the r.m.s.\/
momentum density fluctuations are constant on the horizon scale:
\ben
\int_0^\et\frac{d^3 k}{(2\pi)^3} \vev{|\Th_{0i}(\et,\bk)|^2} \propto
\et^{-4}.
\een
Therefore we may write
\ben
\vev{|\Th_{0i}(\et,\bk)|^2} = V^{-1}\et^{-1} \mu^2 S^{(0i)}(k\et),
\een
where $S^{(0i)}(z)$ is a dimensionless scaling function
associated with the momentum density, and $V$ is the normalisation volume.
Finally, we have
\ben
\vev{|\de(\et)|^2} \to V^{-1}\frac{1}{3}
\(\frac{4\pi G\mu}{10}\)^2 \et^4
\Up k \int_{\et_i}^\et \frac{d\et_1}{\et_1}\,S^{(0i)}(k\et_1).
\een
Thus, provided the momentum scaling function is well-behaved, and the
integral converges, we recover the Harrison-Zel'dovich spectrum.
There is
perhaps a logarithmic correction, if the high frequency limit of the
source power spectrum is white noise, as would be supplied by a
population of small string loops.  Given the expression for $\Th_{0i}$,
easily calculable from (\ref{StrEMTen}), we should expect $S^{(0i)}$ to
be proportional to the mean square string velocity $v^2$ and the
average number of Hubble lengths of string per Hubble volume $\ga^2$.

A similar procedure was adopted by Albrecht
and Stebbins \cite{AlbSte92} to derive an analytic approximation
to the power spectrum for strings.  However, they did not explicitly
remove the initial compensation, and simulated  it
by a window function.  One then finds that the power spectrum
depends on $\vev{|\Th_+|^2}$, which goes as $v^4$. Nevertheless,
the functional form of their approximation that they give is
reasonable, even if the details are not quite right.

In general terms, we learn from the above that the appropriate objects
for the numerical study of the density perturbation power spectrum
are the two-time correlation functions
$\vev{\Th_{00}(\et_1)\Th_{00}^*(\et_2)}$ and
$\vev{\Th_{0i}(\et_1)\Th_{0i}^*(\et_2)}$.  Getting good statistics is
rather hard with fully realistic string simulations \cite{StrSim},
but simulations in flat
space \cite{FSStrSim} run very much faster, and are thus ideal for
numerical experiments on the form of the correlation functions,
and the calculations of the observable quantities \cite{Cou+94}.
It is to be hoped that they will also aid us in understanding
the coarse-grained features
of string evolution and in constructing analytic models
\cite{AusCopKib94}, which attempt to derive dynamical equations
for quantities such as $\ga$ and $v$.

\end{document}